\documentclass[twocolumn,amsmath,amssymb]{revtex4}
\usepackage{graphicx}
\usepackage{bm}
\usepackage{amsmath}
\usepackage{amssymb}
\usepackage{revsymb}

\begin{document}

\title{Thermal near--field radiative transfer between two spheres}

\author{Arvind Narayanaswamy}
\email{arvindn@alum.mit.edu}
\affiliation{%
Department of Mechanical Engineering, Columbia University\\
New York, NY 10027
}%
\author{Gang Chen}%
\email{gchen2@mit.edu}
\affiliation{%
Department of Mechanical Engineering, Massachusetts Institute of Technology\\
Cambridge, MA 02139
}%

\date{\today}

\begin{abstract}
Radiative energy transfer between closely spaced bodies is known to be significantly larger than that predicted by classical radiative transfer because of tunneling due to evanescent waves. Theoretical analysis of near--field radiative transfer is mainly restricted to radiative transfer between two half--spaces or spheres treated in the dipole approximation (very small sphere) or proximity force approximation (radius of sphere much greater than the gap). Sphere--sphere or sphere--plane configurations beyond the dipole approximation or proximity force approximation have not been attempted. In this work, the radiative energy transfer between two adjacent non--overlapping spheres of arbitrary diameters and gaps is analyzed numerically. For spheres of small diameter (compared to the wavelength), the results coincide with the dipole approximation. We see that the proximity force approximation is not valid for spheres with diameters much larger than the gap, even though this approximation is well established for calculating forces. From the numerical results, a regime map is constructed based on two non--dimensional length scales for the validity of different approximations.
\end{abstract}

\maketitle



\section{\label{sec:intro}Introduction}

It is well known that thermal radiative transfer in the near--field is extremely different compared to classical radiative transfer in the far--field. Interference effects and, in particular, near--field effects due to tunneling of evanescent waves play a vital role. When the objects involved in energy transfer can support surface waves, the heat transfer can be enhanced by orders of magnitude compared to far--field values. The near--field exchange between two half--spaces has been well documented in literature \cite{cravalho67a,polder71,loomis94,carminati99a,shchegrov00b,volokitin01a,narayanaswamy03a,volokitin04a}. Theoretical analysis of heat transfer between a sphere and a plane or between two spheres are limited to the sphere being approximated by a point dipole \cite{mulet01b,domingues05a,pendry99,volokitin01a}. In \cite{volokitin01a}, the authors outlined a method capable of dealing with near--field radiative transfer between a sphere and a flat substrate when retardation effects can be neglected. However, numerical difficulties prevented them from obtaining a solution to the problem. It is possible to derive an asymptotic expression for radiative transfer between two large spheres separated by a very small gap (gap is very small compared to the radius of either of the spheres) from the well known results of radiative transfer between two semi--infinite objects \cite{chapuis06a}. This idea is used extensively in determining van der Waals or Casimir force between macroscopic curved objects and is known as the proximity force approximation \cite{hamaker37,derjaguin1956,lamoreaux97a,gies06a} and expressed as the proximity force theorem \cite{blocki77}. The usage of the proximity force approximation to determine near--field radiative transfer between curved surfaces has not been verified by other numerical solutions or by experiments. Generally, the diameter of the sphere involved in the experiments range from a few microns to a few tens of microns, which is no longer in the point dipole approximation. Hence a complete scattering solution to the problem is necessary to verify with experiments as well as gauging the validity of simpler models. In this paper we investigate, for the first time, the near--field radiative heat transfer between two spheres using the dyadic Green's function (DGF) of the vector Helmholtz equation \cite{chew95a,tsang00a,collin90a} and the fluctuation--dissipation theorem \cite{landau69a,rytov59a,rytov87c}. This formalism of fluctuational electrodynamics was pioneered by Rytov \cite{rytov59a} and is used widely for analyzing near--field radiative transfer \cite{polder71,carminati99a,shchegrov00b,volokitin01a,narayanaswamy03a,narayanaswamy05a}. Though the emphasis of this work is on near--field radiative transfer, it should be pointed that this formalism includes far--field contribution to radiative transfer. 

Electromagnetic scattering by a sphere has been very well studied since the seminal work of Mie, almost a century ago. The two sphere scalar and vector scattering problems have also been investigated for almost the same amount of time by many authors \cite{bruning69,bruning71a,bruning71b}. The two sphere problem involves expansion of the field in terms of the vector spherical waves of each of the spheres and re-expansion of the vector spherical waves of one sphere in terms of the vector spherical waves of the second sphere in order to satisfy the boundary conditions. The two sphere problem, and multiple sphere scattering in general, is especially tougher due to the computational demands of determining translation coefficients for the (vector) spherical wave functions \cite{friedman54,stein61a, cruzan62,bruning69}. Recurrence relations for the scalar \cite{chew92a} and vector spherical waves \cite{chew93a} have reduced the computational complexity considerably. In this work, the DGF for the two sphere configuration is determined by satisfying the boundary conditions for fields on the surface of the two spheres. The translation coefficients are determined using the recurrence relations in \cite{chew92a,chew93a}. 

The paper is arranged as follows. In Section \ref{sec:pft}, simplified, asymptotic results for the radiative heat transfer between two spheres of equal radii, based on the dipole approximation and proximity force approximation are presented. In Section \ref{sec:emag} the DGF formulation and fluctuation--dissipation theorem are introduced. The Poynting vector is expressed in terms of the DGF and material properties. In Section \ref{sec:twosph} the two sphere problem is described and the DGF for this configuration is determined in terms of the vector spherical waves of the two spheres. In Section \ref{sec:radflux}, the expression for radiative flux, and thus the spectral conductance, from one sphere to another is determined. Details regarding the convergence of the series solution and numerical solutions for sphere sizes up to 20 $\mu$m in diameter is presented in Section \ref{sec:numres}.

\section{\label{sec:pft} Asymptotic results for near--field thermal radiation}
The purpose of this section is to present an asymptotic results for the radiative heat transfer between two spheres in the dipole limit as well as the proximity force approximation limit. It is possible to define a radiative conductance between the two spheres at temperatures $T_A$ and $T_B$ as:
\begin{equation}
\label{eqn:radcondpft}
G =  \lim_{T_A \rightarrow T_B } \frac{P\left(T_A, T_B\right)}{|T_A-T_B|}
\end{equation}
where $G$ (units WK$^{-1}$) is the radiative conductance, $P\left(T_A, T_B\right)$ is the rate of heat transfer between the two spheres at $T_A$ and $T_B$. It should be noted that $G$ is a function of temperature. If the radii of the two spheres involved in near--field radiative transfer are much smaller than the thermal wavelength ($\lambda_T = \hbar c/k_BT \approx 7.63 \mu$m at 300 K), the spheres can be treated as point dipoles and the conductance between the spheres is given by \cite{volokitin01a,domingues05a}:
\begin{equation}
G\left(T\right) = \frac{3}{4\pi^3d^6} \int_{0}^{\infty} \frac{d\Theta\left(\omega,T\right)}{dT}\alpha_{1}''\left(\omega\right)\alpha_{2}''\left(\omega\right)d\omega
\end{equation}
where $\Theta\left(\omega,T\right) = \hbar \omega/\left(exp\left(\hbar \omega/k_{B}T\right)-1\right)$; $\alpha_{1}''\left(\omega\right)$ and $\alpha_{2}''\left(\omega\right)$ are the imaginary parts of the polarizability of the spheres; $d$ is the center--to--center distance between the spheres. This $d^{-6}$ behavior of conductance between two spheres is valid only when $d \ll \lambda_T$ and $d \gg R_1+R_2 $ \cite{volokitin01a}.

Calculating the forces or heat transfer between two curved objects as the size of the objects increases become computationally difficult. For spheres of large diameters and much smaller gaps, the proximity force approximation is very useful in calculating the forces between the objects and used widely to calculate van der Waals or Casimir forces \cite{hamaker37,derjaguin1956,blocki77,gies06a} with the knowledge of the same forces between two half--spaces. The proximity force approximation was used to calculate the conductance between a sphere and a flat surface \cite{chapuis06a} from the results of the radiative heat transfer between two half--spaces \cite{polder71,mulet02a,narayanaswamy03a,joulain05a}. The conductance per unit area for flat plates is the radiative heat transfer coefficient, $h$ (units Wm$^{-2}$K$^{-1}$). The heat transfer coefficient $h(x)$ in the near--field, especially when dominated by surface polaritons, is known to vary as $1/x^2$, where $x$ is the gap between the half--spaces \cite{pendry99,mulet02}. The spheres are separated by a minimum gap $x$. We will assume that the spheres are of equal radii, $R$. The conductance between two spheres is computed by approximating the spheres to be flat surfaces of varying gap. By doing so we get a relation between $G$ and $h$ given by
\begin{equation}
\label{eqn:htpft}
G(x;R) = \pi x R h(x)
\end{equation}
Since $h(x)$ varies as $1/x^2$, it is expected that $G(x;R)$ varies as $1/x$. A $1/x$ variation of conductance as determined by a more rigorous theory can be taken as evidence of the validity of the proximity type approximation at the value of the gap. Despite their origin in fluctuations of the electromagnetic field, a significant difference between force and flux is that the force decays to a negligible quantity in the far--field whereas thermal flux attains a finite value in the far--field. This implies that those parts of the spheres with larger gaps contribute very little to force whereas they could contribute significanly to flux because of the larger areas involved. Hence a proximity force type approximation would be valid only when the heat transfer is dominated by contributions from the near--field region. Therefore, we expect the result of Eq. \ref{eqn:htpft} to be valid only when $R$ is small enough that near--field radiation dominates and $x/R \rightarrow 0$. The discussion in Section \ref{sec:numres} shows that this is indeed true.
\\
\\
\section{\label{sec:emag}Electromagnetic formulation}
All the materials are assumed to be non--magnetic and defined by a complex, frequency dependent dielectric function, $\varepsilon(\omega)$. To compute the radiative transfer we follow the method pioneered by Rytov \cite{rytov59a,rytov87c} in which the source for radiation is the thermal fluctuations of charges. The Fourier component of the fluctuating electric field, $\bm{E}(\bm{r}_{1}, \omega)$, and magnetic field, $\bm{H}(\bm{r}_{1}, \omega)$, at any point, $\bm{r}_{1}$, outside a volume containing the sources is given by \cite{tsang00a,chew95a}:
\begin{eqnarray}
\label{eqn:EHGreen}
\bm{E}(\bm{r}_{1}, \omega) & = & i\omega \mu_{o} \int_{V} d^{3}r\overline{\overline{\bm{G}}}_{e}(\bm{r}_{1}, \bm{r}, \omega) \cdot \bm{J}(\bm{r}, \omega) \\
\bm{H}(\bm{r}_{1}, \omega) & = & \int_{V} d^{3}r \overline{\overline{\bm{G}}}_{h}(\bm{r}_{1}, \bm{r}, \omega) \cdot \bm{J}(\bm{r}, \omega),
\end{eqnarray}
where $\overline{\overline{\bm{G}}}_{e}(\bm{r}_{1}, \bm{r}, \omega)$ and $\overline{\overline{\bm{G}}}_{h}(\bm{r}_{1}, \bm{r}, \omega)$, the dyadic Green's functions due to a point source at $\bm{r}$, are related by $\overline{\overline{\bm{G}}}_{h}(\bm{r}_{1}, \bm{r},\omega ) = \bm{\nabla_{1}} \times \overline{\overline{\bm{G}}}_{e}(\bm{r}_{1}, \bm{r}, \omega)$; $\bm{J}(\bm{r}, \omega)$ is the Fourier component of the current due to thermal fluctuations; and $\mu_{o}$ is the permeability of vacuum. The integration is performed over the entire volume $V$ containing the source. The DGFs themselves obey the following equations \cite{tsang00a,chew95a}:
\begin{eqnarray}
\label{eqn:elecdyad}
\bm{\nabla} \times \bm{\nabla} \times \overline{\overline{\bm{G}}}_{e}(\bm{r}, \bm{r'}) - \left(\frac{\omega}{c}\right)^{2}\varepsilon(\bm{r})\overline{\overline{\bm{G}}}_{e}(\bm{r}, \bm{r'}) = \overline{\overline{\bm{I}}}\delta(\bm{r}-\bm{r'})
\end{eqnarray}
where $\overline{\overline{\bm{I}}}$ is the identity dyad and $\delta(\bm{r}-\bm{r'})$ is the Dirac--delta function. At the boundary between two dielectric (possibly lossy) materials, the DGF satisfies the following boundary conditions to ensure continuity of tangential electric and magnetic fields:
\begin{equation}
\label{eqn:elecboundA} 
\bm{\hat{n}} \times \overline{\overline{\bm{G}}}_{e}(\bm{r}_1, \bm{r'}) =  \bm{\hat{n}} \times \overline{\overline{\bm{G}}}_{e}(\bm{r}_2, \bm{r'})
\end{equation}
\begin{equation}
\label{eqn:elecboundB}
\bm{\hat{n}} \times \bm{\nabla} \times \overline{\overline{\bm{G}}}_{e}(\bm{r}_1, \bm{r'}) =  \bm{\hat{n}} \times \bm{\nabla} \times \overline{\overline{\bm{G}}}_{e}(\bm{r}_2, \bm{r'})
\end{equation}
where $\bm{r}_1$ and $\bm{r}_2$ are points on either side of the boundary and $\bm{\hat{n}}$ is a unit normal to the boundary surface at $\bm{r}_1$ (or $\bm{r}_2$). In order to compute the spectral Poynting vector at $\bm{r}_{1}$, we must compute the cross spectral density of $E_{i}(\bm{r}_{1}, t)$ and $H_{j}(\bm{r}_{1}, t)$,  $\langle E_{i\omega}H_{j\omega}^{*}\rangle$, where the $*$ denotes the complex conjugate, the brackets denote a statistical ensemble average, and $i$ and $j$ refer to the three cartesian components ($i \neq j$). From Eq. (\ref{eqn:EHGreen}), we can write an expression for $\langle E_{i \omega}H_{j \omega}^{*} \rangle$ as:

\begin{widetext}
\begin{eqnarray}
\label{eqn:EiHj}
 \langle E_{i}(\bm{r}_{1},\omega)H_{j}^{*}(\bm{r}_{1},\omega) \rangle =i\omega \mu_{o} \int_{V} d^{3}r\int_{V} d^{3}r' \{ G_{e_{il}}(\bm{r}_{1},  \bm{r}, \omega)G^{*}_{h_{jm}}(\bm{r}_{1}, \bm{r'}, \omega) \langle J_{l}(\bm{r}, \omega)J^{*}_{m}(\bm{r'}, \omega) \rangle\} & & 
\end{eqnarray} 
\end{widetext}

The fluctuation--dissipation theorem states that the cross spectral density of different components of a fluctuating  current source in equilibrium at a temperature T is given by \cite{landau69a}:
\begin{equation}
\label{eqn:FDT}
\langle J_{l}(\bm{r}, \omega)J^{*}_{m}(\bm{r'}, \omega) \rangle =  
\frac{\epsilon_{o}\epsilon''(\omega) \omega \Theta(\omega, T)}{\pi} \delta_{lm} \delta(\bm{r} - \bm{r'}),
\end{equation}
where $\epsilon''(\omega)$ is the imaginary part of the dielectric function of the source, $\epsilon_{o}$ is the permittivity of vacuum, and $\Theta(\omega, T)$ is given by $\hbar \omega/(exp(\hbar \omega/k_{B}T)-1)$, where $2\pi \hbar$ is Planck's constant and $k_B$ is Boltzmann's constant. Using Eq. (\ref{eqn:EiHj}) and Eq. (\ref{eqn:FDT}), we have

\begin{widetext}
\begin{eqnarray}
\label{eqn:EiHjavg}
\langle E_{i}(\bm{r}_{1},\omega)H_{j}^{*}(\bm{r}_{1},\omega) \rangle = \frac{i\epsilon_{o}\epsilon''(\omega) \mu_o \omega^{2} \Theta(\omega, T)}{\pi} \int_{V} d^{3}r \{(\bm{G}_{e}(\bm{r}_{1},  \bm{r}, \omega).\bm{G}_{h}^{T*}(\bm{r}_{1}, \bm{r}, \omega) ) _{ij} \} & & 
\end{eqnarray}
\end{widetext}
where the superscript $T$ stands for the transpose of the dyad. Once the Green's function for the given configuration is determined, the above integral is computed numerically. Determining the DGF is not a trivial task and the next section is devoted to determining the DGF in the case of the two sphere configuration.

\section{\label{sec:twosph}Two sphere problem}
The configuration of the two spheres is shown in Fig. \ref{fig:twosphgeom}. At the center of each sphere is a coordinate system. Without loss of generality, the two spheres are arranged such that the z--axes of both coordinate systems pass through the line joining the centers. The x--axis (and y--axis) of both systems are parallel to each other so that a given point in space has the same $\phi$ coordinate value in both systems. The two spheres are at temperatures $T_A$ and $T_B$. In order to determine the radiative transfer, we have to determine the DGF when the source point is in the interior of one of the spheres. We shall take the Dirac--delta source point to be in the interior of sphere A. The most convenient way of dealing with DGF in spherical coordinates is to expand the DGF in terms of vector spherical waves \cite{chew95a}, which are solutions of
\begin{equation}
\label{ref:vechelm}
\bm{\nabla} \times \bm{\nabla} \times \bm{P(r)} - k^2\bm{P(r)} = 0
\end{equation}
The vector spherical waves we will need are given by \cite{chew95a}:
\begin{eqnarray}
\label{eqn:Mvecwave}
\bm{M}_{lm}^{(p)}(k\bm{r})& = & z_{l}^{(p)}(kr)\bm{V}_{lm}^{(2)}(\theta,\phi)
\end{eqnarray} 
\begin{eqnarray}
\label{eqn:Nvecwave}
\nonumber \bm{N}_{lm}^{(p)}(k\bm{r})& = & \zeta_{l}^{(p)}(kr)\bm{V}_{lm}^{(3)}(\theta,\phi)+ \\
& &\frac{z_{l}^{(p)}(kr)}{kr}\sqrt{l(l+1)}\bm{V}_{lm}^{(1)}(\theta,\phi)
\end{eqnarray} 
where $\bm{M}_{lm}^{(p)}(k\bm{r})$ and $\bm{N}_{lm}^{(p)}(k\bm{r})$ are vector spherical waves of order $(l,m)$. $l$ can take integer values from $0$ to $\infty$. For each $l$, $|m| \leq l$. The superscript $p$ refers to the radial behavior of the waves. For $p = 1$, the $\bm{M}$ and $\bm{N}$ waves are regular waves and remain finite at the origin and $z_{l}^{(1)}(kr)$
is the spherical bessel function of order $l$. For $p = 3$, the $\bm{M}$ and $\bm{N}$ waves are outgoing spherical waves that are singular at the origin and $z_{l}^{(3)}(kr)$
is the spherical hankel function of the first kind of order $l$. The radial function $\zeta_{l}^{(p)}(x) = \frac{1}{x}\frac{d}{dx}\left(xz_{l}^{(p)}(x)\right)$ .$\bm{V}_{lm}^{(1)}(\theta,\phi)$, $\bm{V}_{lm}^{(2)}(\theta,\phi)$, and $\bm{V}_{lm}^{(3)}(\theta,\phi)$ are vector spherical harmonics of order $(l,m)$. The three vector spherical harmonics can be expressed in terms of the spherical harmonics, $Y_{lm}(\theta,\phi)$ as:
\begin{figure}
\begin{center}
\includegraphics[height=60mm]{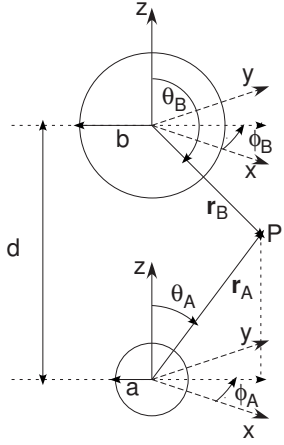}
\caption{\label{fig:twosphgeom} Two sphere configuration. Two non-overlapping spheres of radii a and b are separated by a distance d. At the center of each sphere is a spherical coordinate system oriented such that the two spheres lie along the common z--axis. The x--axes and y--axes are also oriented such that for a given location in space, the $\phi$ coordinate is the same in both coordinate systems. In this figure, the point P has coordinates $(r_A, \theta_A, \phi_A)$ and $(r_B, \theta_B, \phi_B)$ such that $\phi_A = \phi_B$. Region A(B) refers to the interior of sphere A(B). Region C is the exterior of both spheres and is taken to be vacuum.}
\end{center}
\end{figure}
\begin{subequations}
\begin{equation}
\bm{V}_{lm}^{(1)}(\theta,\phi) = \bm{\hat{r}}Y_{lm}
\end{equation}
\begin{equation}
\bm{V}_{lm}^{(2)}(\theta,\phi) = \frac{1}{\sqrt{l(l+1)}}\left(-\bm{\hat{\phi}}\frac{\partial Y_{lm}}{\partial \theta}+ \bm{\hat{\theta}}\frac{im}{sin \theta}Y_{lm}\right)\\
\end{equation}
\begin{equation}
\bm{V}_{lm}^{(3)}(\theta,\phi) = \frac{1}{\sqrt{l(l+1)}}\left(\bm{\hat{\theta}}\frac{\partial Y_{lm}}{\partial \theta}+ \bm{\hat{\phi}}\frac{im}{sin \theta}Y_{lm}\right)
\end{equation}
\end{subequations}
The vector spherical harmonics are orthonormal to each other and satisfy the following relation:
\begin{equation}
\label{eqn:vecsphorth}
\oint_{\Omega} \bm{V}_{lm}^{(r)}(\theta,\phi) \centerdot \bm{V}_{pq}^{(s)*}(\theta,\phi) d\Omega = \delta_{rs}\delta_{lp}\delta_{mq}
\end{equation}
where the integration domain $\Omega$ refers to the surface of a sphere of unit radius and $d\Omega$ is a differential area element on such a sphere. The vector spherical waves $\bm{M}_{lm}^{(p)}(k\bm{r})$ and $\bm{N}_{lm}^{(p)}(k\bm{r})$ are related by $\bm{N}_{lm}^{(p)}(k\bm{r}) = \frac{1}{k}\bm{\nabla} \times \bm{M}_{lm}^{(p)}(k\bm{r})$ and $\bm{M}_{lm}^{(p)}(k\bm{r}) = \frac{1}{k}\bm{\nabla} \times \bm{N}_{lm}^{(p)}(k\bm{r})$. Any solution to Eq. \ref{ref:vechelm} can be expressed as a linear combination of the vector spherical waves. Further properties of spherical harmonics and vector spherical harmonics useful for this analysis is included in Appendix \ref{sec:appvsw} and Appendix A of \cite{narayanaswamy07a}. In this particular case, the field is a linear combination of vector spherical waves of the two coordinate systems as shown in Fig. \ref{fig:twosphgeom}. To satisfy the boundary conditions on the surface of each sphere the vector spherical waves of one coordinate system should be expressed in terms of the vector spherical waves of the other coordinate system. This is what is achieved by means of translation addition theorems for vector spherical waves.

\subsection{\label{sec:vectrans}Coefficients for translation addition theorems}
The vector translation addition theorem \cite{chew90a,chew95a,chew93a} states that:

\begin{subequations}
\label{eqn:vectrans}
\begin{equation}
\begin{split}
\bm{M}_{lm}^{(p)}(k\bm{r}_b) = 
\sum_{\mu = -\nu \atop \nu=1}^{\nu = \infty \atop \mu = \nu} \big[ & A_{\nu\mu}^{lm}(+kd)\bm{M}_{\nu\mu}^{(q)}(k\bm{r}_a) +  \\
& B_{\nu\mu}^{lm}(+kd)\bm{N}_{\nu\mu}^{(q)}(k\bm{r}_a)\big]
\end{split}
\end{equation}
\begin{equation}
\begin{split}
\bm{N}_{lm}^{(p)}(k\bm{r}_b) =
\sum_{\mu = -\nu \atop \nu=1}^{\nu = \infty \atop \mu = \nu}  \big[&B_{\nu\mu}^{lm}(+kd)\bm{M}_{\nu\mu}^{(q)}(k\bm{r}_a) + \\
 &A_{\nu\mu}^{lm}(+kd)\bm{N}_{\nu\mu}^{(q)}(k\bm{r}_a)\big]
\end{split}
\end{equation}
\end{subequations}
The position vectors $\bm{r}_a$ and $\bm{r}_b$ refer to the same location in space in coordinate systems A and B respectively. Computing the coefficients $A_{\nu\mu}^{lm}(+kd)$ and $B_{\nu\mu}^{lm}(+kd)$ has been the topic of many publications \cite{friedman54,stein61a,cruzan62,chew93a}. Generally, the expressions for the coefficients require calculations of Wigner 3j symbols which involve calculations of large number of factorials, making it computationally expensive. Recurrence relations for computing the coefficients efficiently have been proposed by Chew \cite{chew93a}. In the case of the two sphere problem, with translation along the z--axis alone, Eq. \ref{eqn:vectrans} simplifies so that the coefficients are non--zero for $\mu = m$ alone. 

In the region C (exterior to both spheres), the electric and magnetic fields should be expanded in terms of outgoing vector spherical waves of both coordinate systems so that waves decay as $1/r$ as $r \rightarrow \infty$. Hence $\bm{M}_{lm}^{(3)}(k\bm{r}_b)$ and $\bm{N}_{lm}^{(3)}(k\bm{r}_b)$ need to be expressed in terms of $\bm{M}_{\nu m}^{(p)}(k\bm{r}_a)$ and $\bm{N}_{\nu m}^{(p)}(k\bm{r}_a)$ on the surface of sphere A and vice versa. Since $|\bm{r}_a| = a < d$ for all points on the surface of sphere A, only the regular vector spherical waves or $\bm{M}_{\nu m}^{(1)}(k\bm{r}_a)$ and $\bm{N}_{\nu m}^{(1)}(k\bm{r}_a)$ should be used. In addition to $A_{\nu m}^{lm}(+kd)$ and $B_{\nu m}^{lm}(+kd)$, we will also need $A_{\nu m}^{lm}(-kd)$ and $B_{\nu m}^{lm}(-kd)$, which can be obtained through symmetry relations \cite{kim04a}. For further details regarding the computation of the recurrence relations, the reader is referred to \cite{chew92a,chew93a}. 

\subsection{\label{sec:dgf2}DGF - vector spherical wave expansion}
The DGF for any configuration can be split into two parts - one that corresponds to a Dirac--delta source in an infinite medium, $\overline{\overline{\bm{G}}}_{o}$ and one that takes into account the scattering, $\overline{\overline{\bm{G}}}_{sc}$. In this case, the source point is confined to the interior of sphere A. The DGF for source point in sphere A, assuming the whole space to of the same material, is given by:
\begin{widetext}
\begin{equation}
\begin{split}
\label{eqn:homodgf}
\overline{\overline{\bm{G}}}_{o}(\bm{r}_a, \bm{r'}_a) = & \frac{\bm{\hat{r}}\bm{\hat{r}}}{k_a^2}\delta(\bm{r}_a-\bm{r'}_a) + ik_a\sum_{m = -l \atop \l = 1}^{l = \infty \atop m = l}
\begin{cases}
\bm{M}_{lm}^{(1)}(k_a\bm{r}_a)\bm{M}_{l,-m}^{(3)}(k_a\bm{r'}_a) + \bm{N}_{lm}^{(1)}(k_a\bm{r}_a)\bm{N}_{l,-m}^{(3)}(k_a\bm{r'}_a) & \text{ if } r_a < r'_a \\
\bm{M}_{l m}^{(3)}(k_a\bm{r}_a)\bm{M}_{l,-m}^{(1)}(k_a\bm{r'}_a) + \bm{N}_{lm}^{(3)}(k_a\bm{r}_a)\bm{N}_{l,-m}^{(1)}(k_a\bm{r'}_a) & \text{ if } r_a > r'_a
\end{cases} 
\end{split}
\end{equation}
\end{widetext}
In particular, we are interested in the case where $r_a > r'_a$ since the source is inside the sphere A whereas the boundary of interest is the surface of the sphere. The part of the DGF that depends on the boundaries takes different forms in the three regions, A, B, and C. Inside A, the DGF is a combination of $\overline{\overline{\bm{G}}}_{o}$ and $\overline{\overline{\bm{G}}}_{sc}$, whereas outside A the DGF is entirely $\overline{\overline{\bm{G}}}_{sc}$. Each term, $\bm{M}_{l m}^{(3)}(k_a\bm{r}_a)$ or $\bm{N}_{l m}^{(3)}(k_a\bm{r}_a)$, in Eq. \ref{eqn:homodgf} can be thought of as an independent vector spherical waves that produces scattered waves, i.e. coefficients of scattered waves due to $\bm{M}_{lm}^{(3)}(k_a\bm{r}_a)$ (or $\bm{N}_{lm}^{(3)}(k_a\bm{r}_a)$) are completely decoupled from vector spherical waves of other orders. Let us consider the scattered field due to $\bm{M}_{lm}^{(3)}(k_a\bm{r}_a)$. The scattered field in the three regions is given by:
\begin{widetext}
\begin{equation}
\label{eqn:scatfield}
\sum_{\nu=(m,1)}^{\infty}
\begin{cases}
ik_a \big[\big(A^{lM}_{\nu m}\bm{M}_{\nu m}^{(1)}(k_a\bm{r}_a) + A^{lN}_{\nu m}\bm{N}_{\nu m}^{(1)}(k_a\bm{r}_a)\big) + \big(B^{lM}_{\nu m}\bm{M}_{\nu m}^{(3)}(k_a\bm{r}_b) + B^{lN}_{\nu m}\bm{N}_{\nu m}^{(3)}(k_a\bm{r}_b)\big) \big] & \text{ , in A} \\

ik_f \big[\big(C^{lM}_{\nu m}\bm{M}_{\nu m}^{(3)}(k_f\bm{r}_a) + C^{lN}_{\nu m}\bm{N}_{\nu m}^{(3)}(k_f\bm{r}_a)\big) + \big(D^{lM}_{\nu m}\bm{M}_{\nu m}^{(3)}(k_f\bm{r}_b) + D^{lN}_{\nu m}\bm{N}_{\nu m}^{(3)}(k_f\bm{r}_b)\big) \big] & \text{ , in C} \\

ik_b \big[\big(E^{lM}_{\nu m}\bm{M}_{\nu m}^{(3)}(k_b\bm{r}_a) + E^{lN}_{\nu m}\bm{N}_{\nu m}^{(3)}(k_b\bm{r}_a)\big) + \big(F^{lM}_{\nu m}\bm{M}_{\nu m}^{(1)}(k_b\bm{r}_b) + F^{lN}_{\nu m}\bm{N}_{\nu m}^{(1)}(k_b\bm{r}_b)\big) \big] & \text{ , in B}
\end{cases} 
\end{equation}
\end{widetext}
where the symbol $(m,1)$ refers to the greater of $m$ and 1. $A^{lM}_{\nu m}$ is a coefficient of a vector spherical waves of order $(\nu,m)$ that it is produced by a vector spherical waves of order $(l,m)$ . The superscript $M$ ($N$) is to indicate that it is a coefficient of a $\bm{M}$  ($\bm{N}$) wave. In practice, the upper limit for the summation is resticted to a value $N_{m}$ which depends on $k_f d$. The appropriate value of $N_{m}$ will be discussed in Section \ref{sec:numres}. Using Eq. \ref{eqn:scatfield}, Eq. \ref{eqn:elecboundA}, and Eq. \ref{eqn:elecboundB}, the following set of coupled linear equations can be obtained for the coefficients of the vector spherical waves in the scattered field in region C:
\begin{subequations}
\label{eqn:scateqn}
\begin{equation}
\label{eqn:scateqn1}
\begin{split}
C^{lM}_{\eta m}  +  u_{\eta}(a)\sum_{\nu=(m,1)}^{N_{max}}\big[ & D^{lM}_{\nu m} A_{\eta m}^{\nu m}(-k_f d)+ \\ & D^{lN}_{\nu m} B_{\eta m}^{\nu m}(-k_f d)\big] = p^{M}_{\eta}\delta_{\eta l}
\end{split}
\end{equation}
\begin{equation}
\begin{split}
\label{eqn:scateqn2}
C^{lN}_{\eta m} + v_{\eta}(a)\sum_{\nu=(m,1)}^{N_{max}}\big[& D^{lM}_{\nu m} B_{\eta m}^{\nu m}(-k_f d)+ \\ & D^{lN}_{\nu m} A_{\eta m}^{\nu m}(-k_f d)\big] = 0
\end{split}
\end{equation}
\begin{equation}
\begin{split}
\label{eqn:scateqn3}
D^{lM}_{\eta m} + u_{\eta}(b)\sum_{\nu=(m,1)}^{N_{max}}\big[& C^{lM}_{\nu m} A_{\eta m}^{\nu m}(+k_f d)+  \\ & C^{lN}_{\nu m} B_{\eta m}^{\nu m}(+k_f d)\big] = 0
\end{split}
\end{equation}
\begin{equation}
\begin{split}
\label{eqn:scateqn4}
D^{lN}_{\eta m} + v_{\eta}(b)\sum_{\nu=(m,1)}^{N_{max}}\big[& C^{lM}_{\nu m} B_{\eta m}^{\nu m}(+k_f d)+  \\ & C^{lN}_{\nu m} A_{\eta m}^{\nu m}(+k_f d)\big] = 0
\end{split}
\end{equation}
\end{subequations}
where $\eta$ ranges from $(m,1)$ to $N_{m}$. $u_{\eta}(a)$ and $v_{\eta}(a)$ are Mie coefficients that one encounters in the scattering of a spherical wave by a single sphere and are given by
\begin{subequations}
\begin{equation}
\label{eqn:mieu}
u_{\eta}(a) = \frac{k_a\zeta^{(1)}_{\eta}(k_aa)z^{(1)}_{\eta}(k_fa)-k_f\zeta^{(1)}_{\eta}(k_fa)z^{(1)}_{\eta}(k_aa)}{k_a\zeta^{(1)}_{\eta}(k_aa)z^{(3)}_{\eta}(k_fa)-k_f\zeta^{(3)}_{\eta}(k_fa)z^{(1)}_{\eta}(k_aa)}
\end{equation}
\begin{equation}
\label{eqn:miev}
v_{\eta}(a) = \frac{k_a\zeta^{(1)}_{\eta}(k_fa)z^{(1)}_{\eta}(k_aa)-k_f\zeta^{(1)}_{\eta}(k_aa)z^{(1)}_{\eta}(k_fa)}{k_a\zeta^{(3)}_{\eta}(k_fa)z^{(1)}_{\eta}(k_aa)-k_f\zeta^{(1)}_{\eta}(k_aa)z^{(3)}_{\eta}(k_fa)}
\end{equation}
\end{subequations}
Expressions for $u_{\eta}(b)$ and $v_{\eta}(b)$ are obtained by replacing $k_a$ and $a$ by $k_b$ and $b$ respectively. If the original wave is $\bm{N}_{lm}^{(3)}(k_a\bm{r}_a)$ instead of $\bm{M}_{lm}^{(3)}(k_a\bm{r}_a)$, the only difference is that the right hand side (RHS) of Eq. \ref{eqn:scateqn1} becomes 0 and the RHS of Eq. \ref{eqn:scateqn2} becomes $p^{N}_{\eta}\delta_{\eta l}$. $p^{M}_{\eta}$ and $p^{N}_{\eta}$ are given by
\begin{subequations}
\begin{equation}
\label{eqn:miepm}
p^M_{\eta} = \frac{-i/(k_f a)} {k_a a\zeta^{(1)}_{\eta}(k_a a)z^{(3)}_{\eta}(k_fa)-k_f a\zeta^{(3)}_{\eta}(k_fa)z^{(1)}_{\eta}(k_aa)}
\end{equation}
\begin{equation}
\label{eqn:miepn}
p^N_{\eta} = \frac{i/(k_f a)}{k_a a \zeta^{(3)}_{\eta}(k_fa)z^{(1)}_{\eta}(k_aa)-k_f a\zeta^{(1)}_{\eta}(k_aa)z^{(3)}_{\eta}(k_fa)}
\end{equation}
\end{subequations}
\\
For a given $m$, we have $(N_{m}-(m,1) +1)$ $\bm{M}$ waves and $(N_{m} - (m,1) +1)$ $\bm{N}$ waves. We see that the left hand side (LHS) for a given value of $m$ remains the same while the only difference is in the RHS $(p^M_{\eta}$ and $p^N_{\eta})$. Once all the coefficients in Eq. \ref{eqn:scateqn} are obtained, the DGF due to scattering, $\overline{\overline{\bm{G}}}_{sc}$, and its curl, $\bm{\nabla} \times \overline{\overline{\bm{G}}}_{sc}$, can be written as:
\begin{widetext}
\begin{equation}
\label{eqn:scatdgf}
\overline{\overline{\bm{G}}}_{sc}(\bm{r}_a, \bm{r'}_a) = ik_f  \sum_{l,\nu=(1,m) \atop m=-N_{m}}^{m= N_m \atop l,\nu= N_{m}} (-1)^m
\left[\begin{matrix}
\left[
\begin{matrix}
\big(C^{lM}_{\nu m}\bm{M}_{\nu m}^{(3)}(k_f\bm{r}_a) + C^{lN}_{\nu m}\bm{N}_{\nu m}^{(3)}(k_f\bm{r}_a)\big) +\\ 
\big(D^{lM}_{\nu m}\bm{M}_{\nu m}^{(3)}(k_f\bm{r}_b) + D^{lN}_{\nu m}\bm{N}_{\nu m}^{(3)}(k_f\bm{r}_b)\big) 
\end{matrix}\right]
\bm{M}_{l,-m}^{(1)}(k_a\bm{r'}_a) + \\
\left[
\begin{matrix}
\big(C^{'lM}_{\nu m}\bm{M}_{\nu m}^{(3)}(k_f\bm{r}_a) + C^{'lN}_{\nu m}\bm{N}_{\nu m}^{(3)}(k_f\bm{r}_a)\big) +\\ 
\big(D^{'lM}_{\nu m}\bm{M}_{\nu m}^{(3)}(k_f\bm{r}_b) + D^{'lN}_{\nu m}\bm{N}_{\nu m}^{(3)}(k_f\bm{r}_b)\big) 
\end{matrix}\right]\bm{N}_{l,-m}^{(1)}(k_a\bm{r'}_a)
\end{matrix}\right]
\end{equation}

\begin{equation}
\label{eqn:scatdgfmag}
\bm{\nabla} \times \overline{\overline{\bm{G}}}_{sc}(\bm{r}_a, \bm{r'}_a) = ik_f^2 \sum_{l,\nu=(1,m) \atop m=-N_{m}}^{m= N_m \atop l,\nu= N_{m}} (-1)^m
\left[\begin{matrix}
\left[
\begin{matrix}
\big(C^{lM}_{\nu m}\bm{N}_{\nu m}^{(3)}(k_f\bm{r}_a) + C^{lN}_{\nu m}\bm{M}_{\nu m}^{(3)}(k_f\bm{r}_a)\big) +\\ 
\big(D^{lM}_{\nu m}\bm{N}_{\nu m}^{(3)}(k_f\bm{r}_b) + D^{lN}_{\nu m}\bm{M}_{\nu m}^{(3)}(k_f\bm{r}_b)\big) 
\end{matrix}\right]
\bm{M}_{l,-m}^{(1)}(k_a\bm{r'}_a) + \\
\left[
\begin{matrix}
\big(C^{'lM}_{\nu m}\bm{N}_{\nu m}^{(3)}(k_f\bm{r}_a) + C^{'lN}_{\nu m}\bm{M}_{\nu m}^{(3)}(k_f\bm{r}_a)\big) +\\ 
\big(D^{'lM}_{\nu m}\bm{N}_{\nu m}^{(3)}(k_f\bm{r}_b) + D^{'lN}_{\nu m}\bm{M}_{\nu m}^{(3)}(k_f\bm{r}_b)\big) 
\end{matrix}\right]\bm{N}_{l,-m}^{(1)}(k_a\bm{r'}_a)
\end{matrix}\right]
\end{equation}
\end{widetext}

\section{\label{sec:radflux}Radiative flux}
The radiative heat transfer between the two spheres is calculated from the Poynting vector normal to the surface of sphere B, which in turn depends on the tangential fields on the surface of sphere B. The expression for the DGF can be modified to reflect tangential and normal fields on the surface of sphere B by using Eq. \ref{eqn:scateqn3} and  Eq. \ref{eqn:scateqn4}  in Eq. \ref{eqn:scatdgf} and Eq. \ref{eqn:scatdgfmag} and eliminating $ \bm{M}_{\nu m}^{(3)}(k_f\bm{r}_a) $ and $ \bm{N}_{\nu m}^{(3)}(k_f\bm{r}_a)$ to result in the following equations:
\begin{widetext}
\begin{equation}
\label{eqn:scatdgfsim}
\begin{split}
& \overline{\overline{\bm{G}}}_{sc}(\bm{r}_a, \bm{r'}_a) = \\
& = \frac{1}{b} \sum_{l,\nu=(1,m) \atop m  =-N_{m}}^{m= N_m \atop l,\nu= N_{m}} (-1)^m 
 \left[\begin{matrix}
\left[
\begin{matrix}
\left(-\frac{D^{lM}_{\nu m}z_{l}^{(1)}(k_b r_b)}{k_b b\zeta^{(1)}_{\nu}(k_bb)z^{(1)}_{\nu}(k_fb)-k_fb\zeta^{(1)}_{\nu}(k_fb)z^{(1)}_{\nu}(k_bb)}\right)\bm{V}_{\nu m}^{(2)}(\theta_b,\phi_b) & + \\
\left(+\frac{D^{lN}_{\nu m}\zeta_{l}^{(1)}(k_b r_b)}{k_b b\zeta^{(1)}_{\nu}(k_fb)z^{(1)}_{\nu}(k_bb)-k_fb\zeta^{(1)}_{\nu}(k_bb)z^{(1)}_{\nu}(k_fb)}\right)\bm{V}_{\nu m}^{(3)}(\theta_b,\phi_b) & + \\
\left(+\frac{(k_b/k_f)D^{lN}_{\nu m} z_{l}^{(1)}(k_b r_b)(\sqrt{\nu(\nu+1)}/k_bb)}{k_b b\zeta^{(1)}_{\nu}(k_fb)z^{(1)}_{\nu}(k_bb)-k_fb\zeta^{(1)}_{\nu}(k_bb)z^{(1)}_{\nu}(k_fb)}\right)\bm{V}_{\nu m}^{(1)}(\theta_b,\phi_b) &
\end{matrix}\right]
\bm{M}_{l,-m}^{(1)}(k_a\bm{r'}_a) \\
\\
+\left[
\begin{matrix}
\text{similar terms with primed coefficients}
\end{matrix}\right]\bm{N}_{l,-m}^{(1)}(k_a\bm{r'}_a)
\end{matrix}\right]
\end{split}
\end{equation}

\begin{equation}
\label{eqn:scatdgfmagsim}
\begin{split}
& \bm{\nabla} \times \overline{\overline{\bm{G}}}_{sc}(\bm{r}_a, \bm{r'}_a) = \\
& \frac{k_f}{b}\sum_{l,\nu =(1,m) \atop m=-N_{m}}^{m= N_m \atop l,\nu= N_{m}} (-1)^m
\left[\begin{matrix}
\left[
\begin{matrix}
\left(-\frac{(k_b/k_f)D^{lM}_{\nu m}\zeta_{l}^{(1)}(k_b r_b)}{k_b b\zeta^{(1)}_{\nu}(k_bb)z^{(1)}_{\nu}(k_fb)-k_fb\zeta^{(1)}_{\nu}(k_fb)z^{(1)}_{\nu}(k_bb)}\right)\bm{V}_{\nu m}^{(3)}(\theta_b,\phi_b) & + \\
\left(+\frac{(k_b/k_f)D^{lN}_{\nu m}z_{l}^{(1)}(k_b r_b)}{k_b b\zeta^{(1)}_{\nu}(k_fb)z^{(1)}_{\nu}(k_bb)-k_fb\zeta^{(1)}_{\nu}(k_bb)z^{(1)}_{\nu}(k_fb)}\right)\bm{V}_{\nu m}^{(2)}(\theta_b,\phi_b) & - \\
\left(+\frac{D^{lN}_{\nu m} z_{l}^{(1)}(k_b r_b)(\sqrt{\nu(\nu+1)}/k_bb)}{k_b b\zeta^{(1)}_{\nu}(k_bb)z^{(1)}_{\nu}(k_fb)-k_fb\zeta^{(1)}_{\nu}(k_fb)z^{(1)}_{\nu}(k_bb)}\right)\bm{V}_{\nu m}^{(1)}(\theta_b,\phi_b) &
\end{matrix}\right]
\bm{M}_{l,-m}^{(1)}(k_a\bm{r'}_a) \\
\\
+\left[
\begin{matrix}
\text{similar terms with primed coefficients}
\end{matrix}\right]\bm{N}_{l,-m}^{(1)}(k_a\bm{r'}_a)
\end{matrix}\right]
\end{split}
\end{equation}
\end{widetext}
Using Eq. \ref{eqn:EiHjavg}, Eq. \ref{eqn:scatdgfsim}, Eq. \ref{eqn:scatdgfmagsim}, Eq. \ref{eqn:veccross23}, Eq. \ref{eqn:veccross32}, Eq. \ref{eqn:MMsource}, Eq. \ref{eqn:NNsource}, and some algebraic manipulation to yield this expression for the spectral radiative transfer between the two spheres, one at temperature $T_A$ and the other at $T_B$:
\begin{widetext}
\begin{equation}
\label{eqn:radtrans}
\begin{split}
& P\left(\omega; T_A, T_B\right) = (\Theta(\omega,T_A)-\Theta(\omega,T_B))\frac{a}{b} \times  \\
& \times \sum_{m, l, \beta}
\left[\begin{matrix}
\left[\Im \left(\frac{1}{x_{_\beta}(b)} \right) \left|\frac{z^{(1)}_{l}(k_a a) D_{\beta m}^{l M}}{z^{(1)}_{\beta}(k_fb)} \right|^2 - \Im \left(\frac{1}{y_{_\beta}(b)} \right) \left|\frac{z^{(1)}_{l}(k_a a) D_{\beta m}^{l N}}{r^{(1)}_{\beta}(k_fb)} \right|^2\right]\Im \left(\frac{1}{x_{_l}(a)} \right) |x_{_l}(b)|^2 + \\
\left[\Im \left(\frac{1}{x_{_\beta}(b)} \right) \left|\frac{r^{(1)}_{l}(k_a a) D_{\beta m}^{'l M}}{z^{(1)}_{\beta}(k_fb)} \right|^2 - \Im \left(\frac{1}{y_{_\beta}(b)} \right) \left|\frac{r^{(1)}_{l}(k_fa) D_{\beta m}^{'l N}}{r^{(1)}_{\beta}(k_fb)} \right|^2\right]\Im \left(\frac{1}{y_{_l}(a)} \right) |y_{_l}(b)|^2
\end{matrix}\right]
\end{split}
\end{equation}
\end{widetext}
where $x_{_l}(a) = k_a a \zeta^{(1)}_{\eta}(k_aa)z^{(1)}_{\eta}(k_fa)-k_f a \zeta^{(1)}_{\eta}(k_fa)z^{(1)}_{\eta}(k_aa)$ and $y_{_l}(a) = k_a a \zeta^{(1)}_{\eta}(k_fa)z^{(1)}_{\eta}(k_aa)-k_f a \zeta^{(1)}_{\eta}(k_aa)z^{(1)}_{\eta}(k_fa)$. Just as in Eq. \ref{eqn:radcondpft}, it is possible to define a spectral radiative conductance between the two spheres at a temperature $T$ ( $T_A \rightarrow T,T_B \rightarrow T $) as:
\begin{widetext}
\begin{equation}
\label{eqn:speccond}
\begin{split}
G\left(\omega;T \right) & = \lim_{T_A \rightarrow T_B} \frac{P\left(\omega; T_A, T_B\right)}{|T_A-T_B|} = k_B  \frac{X^2 e^X}{\left(e^X - 1\right)^2} \frac{a}{b} \times \\
& \sum_{m, l, \beta}
\left[\begin{matrix}
\left[\Im \left(\frac{1}{x_{_\beta}(b)} \right) \left|\frac{z^{(1)}_{l}(k_a a) D_{\beta m}^{l M}}{z^{(1)}_{\beta}(k_fb)} \right|^2 - \Im \left(\frac{1}{y_{_\beta}(b)} \right) \left|\frac{z^{(1)}_{l}(k_a a) D_{\beta m}^{l N}}{r^{(1)}_{\beta}(k_fb)} \right|^2\right]\Im \left(\frac{1}{x_{_l}(a)} \right) |x_{_l}(b)|^2 +\\
\left[\Im \left(\frac{1}{x_{_\beta}(b)} \right) \left|\frac{r^{(1)}_{l}(k_a a) D_{\beta m}^{'l M}}{z^{(1)}_{\beta}(k_fb)} \right|^2 - \Im \left(\frac{1}{y_{_\beta}(b)} \right) \left|\frac{r^{(1)}_{l}(k_fa) D_{\beta m}^{'l N}}{r^{(1)}_{\beta}(k_fb)} \right|^2\right]\Im \left(\frac{1}{y_{_l}(a)} \right) |y_{_l}(b)|^2
\end{matrix}\right]
\end{split}
\end{equation}
\end{widetext}
where $X = \hbar \omega/k_B T$.
It can be seen from Eq. \ref{eqn:speccond} that the spectral conductance has units of $k_B (JK^{-1})$ and can split into two parts, one that depends on temperature and the other that is obtained from the DGF of the two sphere problem. The radiative conductance between the two particles that one would measure in an experiment is the integral of $G\left(\omega;T \right)$. 
\begin{equation}
\label{eqn:totalcond}
G_t(T) = \int\limits_{0}^{\infty} G\left(\omega;T \right) d\omega
\end{equation}

\section{\label{sec:numres}Numerical results}
To ensure that the program written to determine the near--field radiative transfer heat transfer is not misbehaving, a few tests can be performed. One of them is agreement between the numerical results and the analytic expression for conductance in the point dipole limit. For spherical particles in the 1 nm to 50 nm radius, the numerical results agree well with the expression for conductivity in the point dipole limit \cite{mulet01a,domingues05a}. In addition to this another test to ensure the correctness of the method is based on the principle of detailed balance. The radiative conductance between two spheres of arbitrary radii must be independent of the numbering scheme for naming the particles, i.e $G_{12} = G_{21}$, where the subscripts 1 and 2 refer to the two spheres. This is necessary to ensure that when the two spheres are at the same temperature the net heat transfer between the two spheres is zero. It is indeed seen from results shown in Table \ref{table:conductancerecip} that by switching the position if the spheres, keeping the gap the same, results in the same value of conductance (the relative errors are generally of the order 10$^{-14}$)

\begin{table}[h]
\caption{\label{table:conductancerecip}Conductance obtained for spheres of unequal radii. By swapping the radii of the spheres, it is seen that the value of conductance remains the same.}
\begin{center}
\begin{tabular}{|c|c|c|c|}
\hline
Gap ($\mu$m) & a ($\mu$m)& b ($\mu$m)& Conductance (WK$^{-1}$)\\ \hline
0.5	&	1 	&	2  & 	1.63816 $\times 10^{-11}$\\ 
0.5	&	2 	&	1  & 	1.63816 $\times 10^{-11}$\\ 
0.8	&	2 	&	3  &	3.34409 $\times 10^{-11}$\\ 
0.8	&	3 	&	2  &	3.34409 $\times 10^{-11}$\\ \hline
\end{tabular}
\end{center}
\end{table}
\subsection{\label{subsec:numconv}Convergence analysis}
Though the two-sphere scattering problem has been discussed in literature, the near--field interaction between the two spheres has not been analyzed in detail. In particular, the number of terms required for convergence for the near--field problem has not been mentioned at all. For Mie scattering by a single sphere of radius $a$ and wavevector magnitude $k$ the number of terms for convergence, $N_{conv}$, is given by \cite{bruning69}:
\begin{equation}
\label{eqn:mieconvff}
N_{conv} = 1+ka+3(ka)^{1/3}
\end{equation}
For the two sphere problem a slightly different criterion based on the center to center distance between the two spheres is proposed and given by \cite{gumerov02a}:
\begin{equation}
\label{eqn:twosphff}
N_{conv} = \frac{1}{2}e kd
\end{equation}
where $e$ is the base of the natural logarithm. Both Eq. \ref{eqn:mieconvff} and Eq. \ref{eqn:twosphff} are valid criteria for computing far--field quatities, like the scattering coefficient. For the near--field problem, we expect the gap between the two spheres to be of great importance. To determine the number of terms required for convergence of near--field quantities we seek parallels to the much simpler and often investigated problem of near--field transfer between two half--spaces. In the near--field two half--space problem, the equivalent of the number of terms for convergence is the truncation for the in--plane wave vector. For a gap $x$ between the two half--spaces, the predominant contribution to radiative flux in those frequency regions where electromagnetic surface waves are important is from in--plane wave vectors up to the order of $k_{in} \approx 1/x$. This is true when $\sqrt{|\varepsilon|} \approx 1$, where $\varepsilon$ is the dielectric function of the material of the sphere. This dielectric function of silica between the frequencies of 0.04 eV to 0.16 eV satisfies this condition, in which range the maximum value of $\sqrt{|\varepsilon|}$ is $\approx 2.1$. If we can draw an analogy between the in--plane wave vectors in the two half--space problem and the two sphere problem, we can propose a criterion for convergence. In the case of the two-sphere problem, the in--plane wave vector equivalent is given by the wavelength of periodic variations on the surface of the sphere. A given vector spherical harmonic, $\bm{V}_{lm}^{(p)}(\theta,\phi)$, $(p = 1,2,\text{or } 3)$ corresponds to a variation $\exp(im\phi)$ along the equator of the sphere. The period of this particular vector spherical harmonic (along the equator) is $2\pi a/m$ and the corresponding wave vector is $m/a$, where $a$ is the radius of the sphere. For a given gap $x$ between two spheres of radii $a$ (equal for now), the number of terms $N_{conv}$ for convergence should be chosen such that $(N_{conv}/a)x \approx 1$. Hence the convergence criterion for near--field effects is given by:
\begin{equation}
\label{eqn:twosphnf}
N_{conv} \approx \frac{a}{x}
\end{equation}
For spheres of unequal radii the number of terms depends on the greater of the two radii. The convergence criterion proposed here is an upper limit and depending on the optical properties of the spheres, it could be considerably lesser. For spheres which exhibit surface wave resonances, Eq. \ref{eqn:twosphnf} is a valid convergence criterion as shown by our numerical investigations. Depending on the configuration of the spheres, the criterion for convergence is given by $max(\frac{1}{2}e kd, \frac{a}{x})$.

\begin{figure}[h]
\begin{center}
\includegraphics[height=60mm, bb=42 163 538 568]{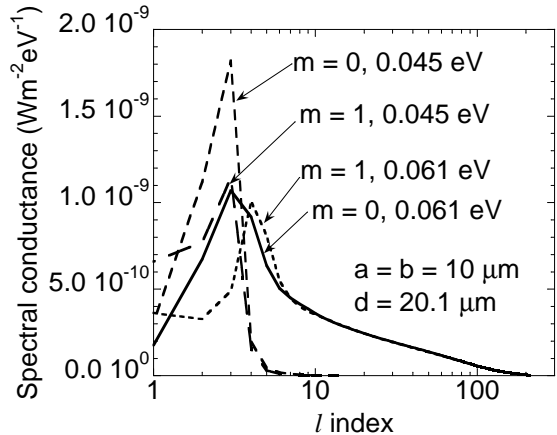} 
\caption{\label{fig:nmaxconvres} Plotted in this figure is the contribution to spectral conductance as a function of $l$ for $m = 0, 1$ at 0.061 eV and 0.045 eV for two spheres of equal radii $a = b = 10$ $\mu$m at a gap of 100 nm ($d = 20.1 $ $\mu$m). Surface phonon polaritons contribute significantly to the radiative transfer at 0.061 eV and not at 0.045 eV.  Hence the number of terms required for convergence is significantly lesser than that prescribed by Eq. \ref{eqn:twosphnf}}
\end{center}
\end{figure}

Since silica spheres are easily available for experimental investigation, we will present our numerical results of radiative transfer between two silica spheres of equal radii at 300 K. In addition, silica is a polar material and hence supports surface phonon polaritons in certain frequency ranges. The dielectric function of silica is taken from \cite{palik1}. The real part of the dielectric function is negative for silica in two frequency ranges in the IR -- from 0.055 to 0.07 eV and 0.14 to 0.16 eV. It is expected that (and shown later) that surface phonon polariton resonances occur in the these frequency ranges. To confirm the prediction of Eq. \ref{eqn:twosphnf}, the contribution to the spectral conductance from each $(l,m)$ mode is plotted as a function of $l$ for different values of $m$ in Fig. \ref{fig:nmaxconvres}. The spectral conductance is plotted at two frequencies -- 0.061 eV and 0.045 eV. The surface polariton mode exists at 0.061 eV but not at 0.045 eV. 

\begin{figure}[h]
\begin{center}
\includegraphics[height=60mm, bb=45 160 558 568]{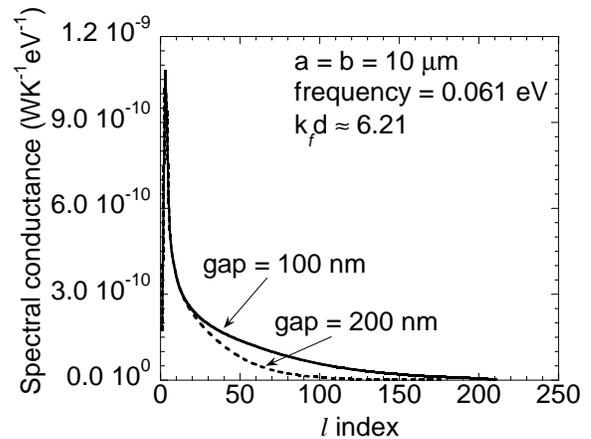} 
\caption{\label{fig:3}  Plot of spectral conductance at 0.061 eV between two spheres with equal radii of 10 $\mu$m at gaps of 100 nm and 200 nm. The curves plotted are for m = 0.  }
\end{center}
\end{figure}

In Fig. \ref{fig:3} the spectral conductance at 0.061 eV for $m$ = 0 between two spheres of radius 10 $\mu$m is plotted as a function of $l$ for two different gaps -- 100 nm and 200 nm. As the gap doubles, the number of terms required for convergence approximately reduces by half, confirming Eq. \ref{eqn:twosphnf}. For instance, the spectral conductance is $1.02 \times 10^{-11} \text{WK}^{-1}\text{eV}^{-1}$ at $l = 181$ for 100 nm gap and $l = 99$ for 200 nm gap.
The contribution to spectral conductance from smaller values of $l$ ($l \lessapprox 15$), corresponding to propagating waves, does not change much with variation in gap. It is the contribution from larger values of $l$, corresponding to surface waves, that changes appreciably with gap. Based on these results we use at least $N_{max} = 2a/(d-2a)$ terms in our computations. Because of the computation difficulties in solving Eq. \ref{eqn:scateqn}, we present results for a maximum radius to gap ratio of 100 for computations at one frequency and 100 frequency points (unequally spaced, with greater density in the resonant parts of the spectrum) over the spectrum from 0.04 eV to 0.16 eV. Equation \ref{eqn:scateqn} is solved using the software package Mathematica.

\begin{figure}[h]
\begin{center} 
\includegraphics[height = 60mm, bb=54 163 513 567]{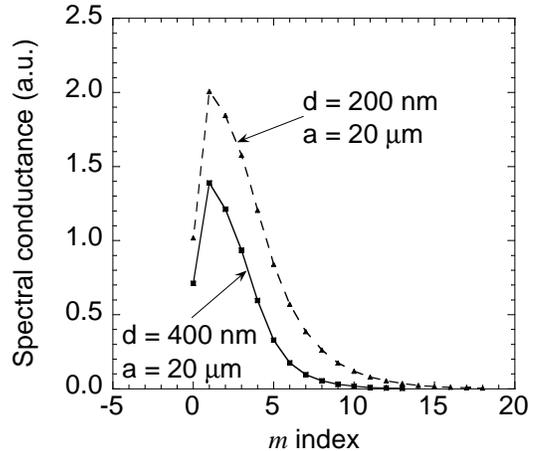} 
\caption{\label{fig:convvsm} Contribution to spectral conductance from each value of $m$ at 0.061 eV. The curves shown are for spheres of radius 20 $\mu$m and gaps of 200 nm and 400 nm between them.}
\end{center}
\end{figure}

As mentioned earlier, the symmetry associated with the two-sphere problem allows for solving for the contribution from each value of the $m$ independently, starting from $m = 0$ and proceeding with increasing values of $m$. As $m$ increases, the contribution to conductance decreases as shown in Fig. \ref{fig:convvsm}. The computation is stopped when a vacule of $m$ is reached such that the contribution to conductance is less than 5 $\times$ 10$^{-3}$ times the contribution from $m = 0$. Even though the contribution to conductance is significant for terms with $l \approx N_{max}$, the contributions from $m$ drops much faster. This is fortunate - the time taken to compute the conductance is proportional to the number of values of $m$ required.
\\

\subsection{\label{subsec:speccond}Spectral conductance}
\begin{figure}[h]
\begin{center}
\includegraphics[height = 60mm, bb=54 163 513 567]{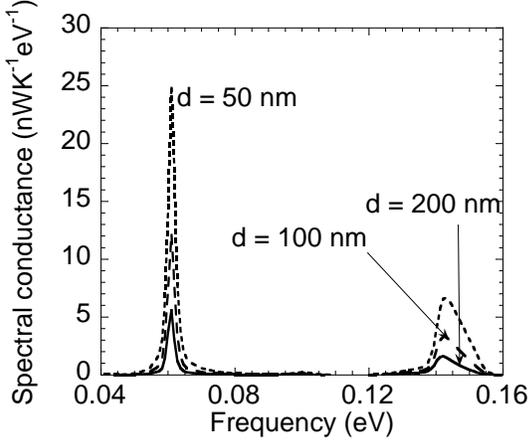} 
\caption{\label{fig:spectralcond1mic} Plot of spectral conductance between two silica spheres of 1 $\mu$m radii at gaps of 50 nm, 100 nm, and 200 nm from 0.04 eV to 0.16 eV. Surface phonon polaritons in the 0.055 to 0.07 eV range and in the 0.14 to 0.16 eV range contribute to the two peaks seen in the figure. }
\end{center}
\end{figure}

Unlike the case of near--field radiative transfer between two half--spaces, where the conductance is a function of only the gap between the half--spaces (and the optical properties of the half--spaces and interveing medium), the conductance in the case of sphere--to--sphere radiative transfer varies as a function of the gap as well as the size of the sphere. A gradual transition occurs from a region of near--field dominated radiation to that of far--field dominated radiation. In Fig. \ref{fig:spectralcond1mic} the spectral conductance between two silica spheres of 1 $\mu$m radius is plotted as a function of frequency. We see from the two peaks that the conductance is dominated by the surface phonon polariton regions. 

\begin{figure}[h]
\begin{center} 
\includegraphics[height = 60mm, bb=54 163 513 567]{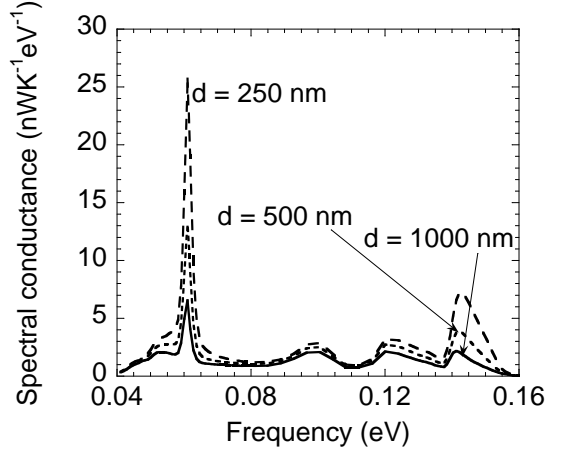} 
\caption{\label{fig:spectralcond5mic} Plot of spectral conductance between two silica spheres of 5 $\mu$m radii at gaps of 250 nm, 500 nm, and 1000 nm from 0.04 eV to 0.16 eV. The gaps have been chosen so as to maintain the same ratio of gap to radius as for the curves shown in Fig. \ref{fig:spectralcond1mic}.}
\end{center}
\end{figure}

\begin{figure}[h]
\begin{center} 
\includegraphics[height = 60mm, bb=54 163 513 567]{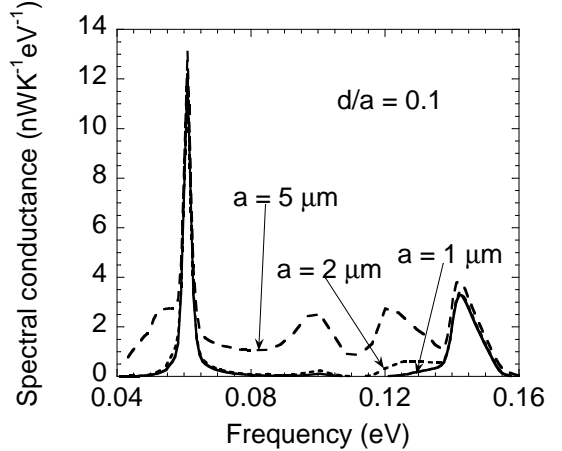} 
\caption{\label{fig:speccondsamedbyRratio}  Spectral conductance of spheres of radii 1 $\mu$m, 2 $\mu$m, and 5 $\mu$m at gaps of 100 nm, 200 nm, and 500 nm respectively. The value of gap to radius for all the curves is 0.1.}
\end{center}
\end{figure}

\begin{figure}[h]
\begin{center} 
\includegraphics[height = 60mm, bb=54 163 513 567]{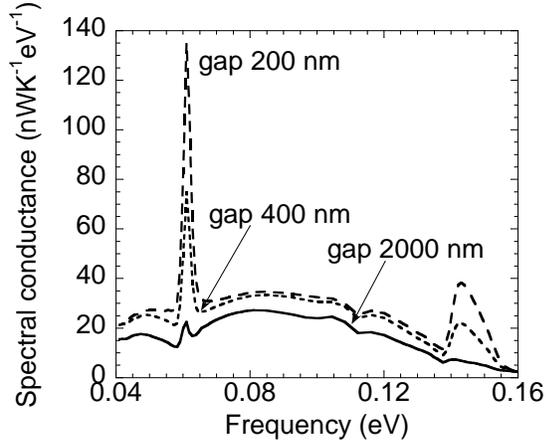} 
\caption{\label{fig:spectralcond20mic}Plot of spectral conductance between two silica spheres of 20 $\mu$m radii at gaps of 200 nm, 400 nm, and 2000 nm from 0.04 eV to 0.16 eV. Unlike the case of plane-to-plane near--field radiative heat transfer, where the contribution from surface polaritons dominate, the conductance between the two spheres has comparable contributions from the resonant and non--resonant regions.  }
\end{center}
\end{figure}

\begin{figure}[h]
\begin{center} 
\includegraphics[height = 60mm, bb=54 163 513 567]{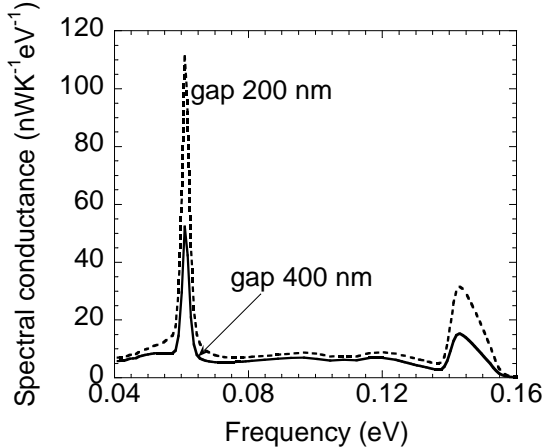} 
\caption{\label{fig:spectralcond20micdiff} The increase in spectral conductance from 2000 nm gap in Fig. \ref{fig:spectralcond20mic} is shown here for gaps of 200 nm and 400 nm. }
\end{center}
\end{figure}

The spectral conductance between the two spheres for larger radii, shown in Fig. \ref{fig:spectralcond5mic}, displays several features of interest. The ratio of radius to gap is maintained the same as in Fig. \ref{fig:spectralcond1mic}. Though the height of the peaks remain approximately the same in this figure as well as Fig. \ref{fig:spectralcond1mic}, the significant difference is from the contribution from those frequency regions that do not support surface polaritons (0.04--0.055 eV, 0.07--0.14 eV). The contribution to the conductance from these ranges do not vary significantly with gap (as long as gap $\ll$ radius). The relation between the ratio of gap to radius and the contribution to spectral conductance is also illustrated in Fig. \ref{fig:speccondsamedbyRratio}. In Fig. \ref{fig:speccondsamedbyRratio}, the spectral conductance of spheres of radii 1 $\mu$m, 2 $\mu$m, and 5 $\mu$m at gaps of 100 nm, 200 nm, and 500 nm respectively are shown. Since the ratio of gap to radius is a constant (0.1), we expect from the asymptotic theory that value of spectral conductance should also be the same in all three cases. We see from the data that this is approximately the case in the regions where electromagnetic surface waves dominate the heat transfer. In the rest of the region, where near--field radiative transfer is non--resonant, increasing radius leads to increased contribution from propagating waves. The spectral conductance as the radius of the spheres is increased to 20 $\mu$m is shown in Fig. \ref{fig:spectralcond20mic}. The increased contribution from the non--resonant parts of the spectrum is evident from the graph. This has an important implication from an experimental point of view. We should be careful not to increase the size of the sphere to such an extent that the resonant radiative transfer is swamped by the non--resonant radiative transfer. For a 20 $\mu$m sphere, we can see that the increase in conductance as the gap decreases is still predominantly due to electromagnetic surface waves as shown in Fig. \ref{fig:spectralcond20micdiff}. In Fig. \ref{fig:spectralcond20micdiff}, the increase in spectral conductance as the gap is decreased from 2000 nm is plotted. Compared to Fig. \ref{fig:spectralcond20mic}, the signature of electromagnetic surface waves is clearer from this plot.  

\begin{figure}[h]
\begin{center} 
\includegraphics[height = 60mm, bb=45 176 525 583]{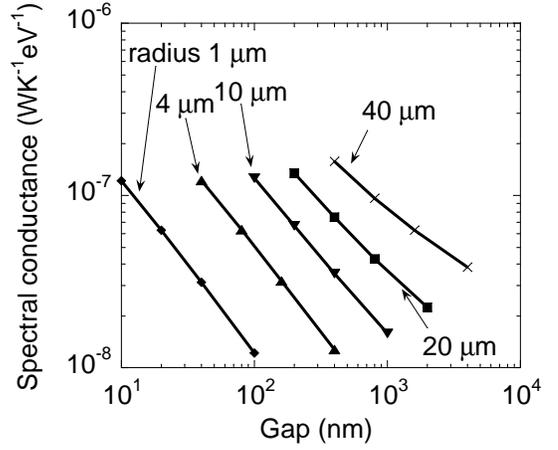} 
\caption{\label{fig:speccondat61mev} Plot of spectral conductance between two spheres of equal radii at 0.061 eV as a function of gap for various radii. For each of the curves, the spectral conductance is computed for the same values of radius/gap. The markers on each curves corresponds to radius/gap values of 100, 50, 25, and 10.}
\end{center}
\end{figure}

In Fig. \ref{fig:speccondat61mev} the spectral conductance at 0.061 eV (corresponding to the first peak in Fig. \ref{fig:spectralcond1mic}) between two spheres is plotted as a function of gap for different values of the radii. The exponent of a power law fit (of the form $y = Ax^B$) to the data points in Fig. \ref{fig:speccondat61mev} is -1.001, -0.984, -0.9024, -0.7781,-0.6119 for radius 1 $\mu$m, 4 $\mu$m, 10 $\mu$m, 20 $\mu$m, and 40 $\mu$m respectively. As the radius increases from 1 $\mu$m to 40 $\mu$m, the slope of the curve decreases, indicating an increased contribution from propagating waves. The behavior at smaller radii can be predicted from the variation of conductance with gap between two planes and the proximity force approximation, as discussed in Section \ref{sec:pft}. For larger diameters, the proximity force type approximation is seen to fail.

\subsection{Total conductance}
\begin{figure}
\begin{center} 
\includegraphics[height = 60mm, bb=54 163 513 567]{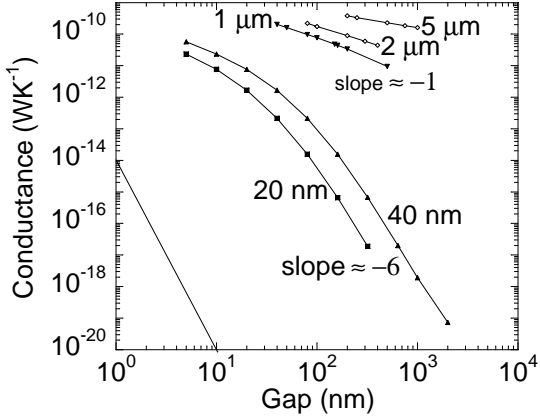} 
\caption{\label{fig:totcond5mic} Total conductance plotted as a function of gap. The number next to each curve is the corresponding radius of the spheres. For 20 nm and 40 nm spheres, the slope of the conductance vs gap curve is approximately -6 for values of gap larger than the radius of spheres. As the radius of the spheres is increased, the slope gradually changes to approximately -1, corresponding to the asymptotic theory.}
\end{center}
\end{figure}

\begin{figure}
\begin{center} 
\includegraphics[height = 60mm, bb=54 163 513 567]{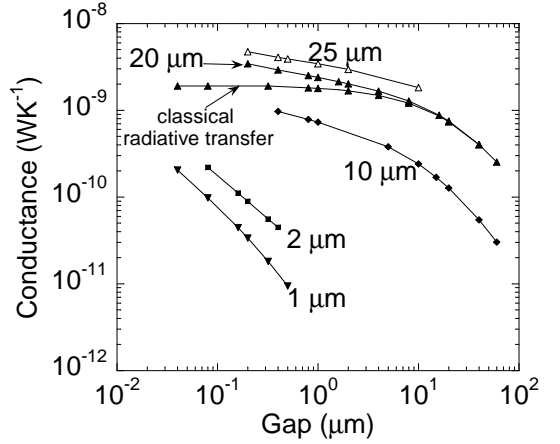} 
\caption{\label{fig:totcondlargedia} Variation of total conductance with gap for various sphere radii. }
\end{center}
\end{figure}

\begin{figure}
\begin{center} 
\includegraphics[height = 60mm, bb=54 163 513 567]{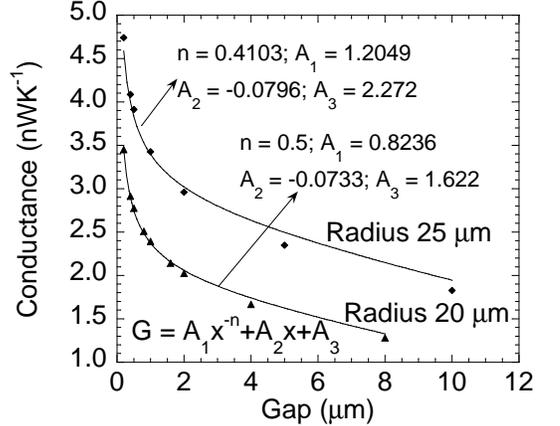} 
\caption{\label{fig:totcond10mic20mic} Variation of total conductance for spheres of radii 20 $\mu$m and 25 $\mu$m with gap. The computation has been restricted to a minimum gap of 200 nm because of the stringent numerical requirements of convergence discussed earlier. Notice that the plot is no longer on a log--log scale.}
\end{center}
\end{figure}

The frequency limits for the calculation of conductance is taken to be 0.041 eV to 0.164 eV. The contribution to conductance from the rest of the frequency spectrum at 300 K is not significant for spheres of smaller radii. However, it is seen from Fig. \ref{fig:spectralcond20mic} that frequencies below 0.04 eV contribute to the non--resonant heat transfer. This will not affect the increase in radiative transfer as the gap is decreased because that increase comes predominantly from the regions supporting surface waves. The total conductance for spheres up to radius of 5 $\mu$m are plotted against gap in Fig. \ref{fig:totcond5mic}. The slope of -6 (approximate) for spheres of radius 20 nm and 40 nm, which can be approximated as point dipoles when the gap between the spheres is much larger than the radius, is in correspondence with the results from the dipole approximation. However, the dipole theory predicts that the conductance should flatten and reach a finite value as the gap decreases to zero, i.e., a slope of zero. What happens in fact is that the near--field effects begin contributing as the gap decreases and the slope in fact decreases from -6 to approximately -1, once again corresponding to the asymptotic theory. However, as the radius increases to 5 $\mu$m, the slope decreases further. This decrease is because of increased contribution from non--resonant regions of the spectrum, as seen in Fig. \ref{fig:spectralcond5mic}. For larger values of spheres, the conductance is plotted in Fig. \ref{fig:totcondlargedia}. On a log-log scale as plotted in Fig. \ref{fig:totcondlargedia}, the near--field effects are apparent for the spheres of 1 $\mu$m and 2 $\mu$m. The reason it does not seem so for the spheres of larger diameter is because of the large contribution from the non--resonant parts of the spectrum, as evidenced from the curve corresponding to the results of classical radiative transfer for the 20 $\mu$m sphere. To understand the effects of near--field transfer for larger spheres, the total conductance for 20 $\mu$m and 25 $\mu$m spheres is plotted in Fig. \ref{fig:totcond10mic20mic}. The data points are fit with a curve $G = A_1x^{-n} + A_2x + A_3$, where $x$ is the gap, $A_1x^{-n}$ is the near--field contribution, $A_2x$ is the contribution due to non--resonant parts of the spectrum (as well as any changes from classical effects due to increase in view factor). We see that the value of the exponent $n$ is 0.5574 for the 20 $\mu$m sphere and 0.5035 for the 25 $\mu$m sphere. If the asymptotic theory we valid at these values of the gap, one would have expected a value of $n = 1$. It is expected that as the gap decreases, the conductance will approach the form predicted in Eq. \ref{eqn:radcondpft} by the asymptotic theory (not confirmed in this paper due to computational restrictions). 

\section{Regime map for the two-sphere problem}
\begin{figure}
\begin{center} 
\includegraphics[height = 60mm, bb=35 195 560 602]{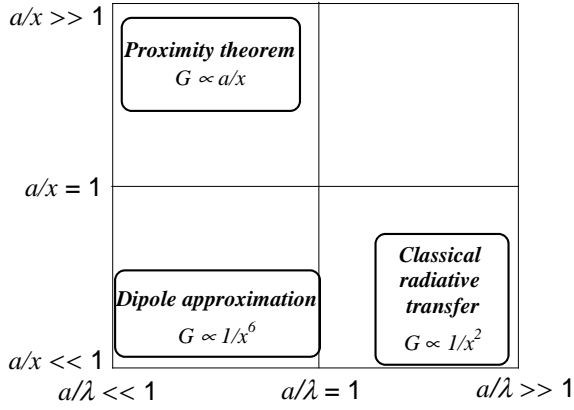} 
\caption{\label{fig:regimemap} Regime map for the two sphere problem. Radius of spheres is $a$, the gap between them is $x$, and the wavelength of radiation is $\lambda$. The numerical solution to the two sphere problem is valid everywhere, limited only by computational restrictions.}
\end{center}
\end{figure}

Unlike the two half--space problem \cite{cravalho67a,polder71,loomis94,carminati99a,shchegrov00b,volokitin01a,narayanaswamy03a,volokitin04a}, the two sphere problem, when other relavant length scales such as the skin depth are unimportant or $|\sqrt{\varepsilon}| \approx 1$ (as in the case of silica spheres), where $\varepsilon$ is the dielectric function of the sphere, has three length scales - the wavelength in consideration $\lambda$, the radius of the spheres $a$, and the gap between the spheres, $x$. Depending on the ratio of the length scales, different approximate theories can be used to predict the conductances. These regions in which different theories are applicable can be represented on a regime map with the two axes representing two non--dimensional length scales, as shown in Fig. \ref{fig:regimemap} and discussed below.
\begin{enumerate}
\item $a \gg \lambda$ and $x > \lambda$: In this case classical radiative transfer can be employed. However, while near--field effects may not be important, interference effects can become important. When $a \approx \lambda$, diffraction effects prevent the usage of classical radiative transfer for even emission from a single sphere.
\item $a \ll \lambda$ and $a \ll x \ll \lambda$:  When the dipole moment of the particles is the dominant contributor to the radiative transfer, the point dipole approximation can be used. However, it should be mentioned that even though the gaps are numerically small, this is not a near--field effect in the sense discussed in this paper.
\item $x \ll a < \lambda$: In this case, the conductance is sees to vary linearly with $1/x$ and is indicative of the validity of the proximity approximation in this regime.
\end{enumerate}
The numerical solution to the two sphere problem as discussed in this paper is valid in all regions of the map, limited only by computational demands.  

\section{\label{sec:conc}Conclusion}
Near-field radiative transfer between two spheres has been investigated for the very first time using fluctuational electrodynamics formalism. Numerical results for spheres of equal radii have been presented and analyzed. It is seen that the proximity force approximation type theory for near--field radiative transfer is valid only when the radius and gap satisfy the condition $x \ll a \leq \lambda$. For larger spheres, the proximity force approximation, which is widely used in calculating forces, is not valid for the gaps analyzed in this paper. The purpose of solving the two sphere problem is to extend the theoretical and numerical formulations of near--field radiative transfer to configurations of objects which can be tested experimentally. The solution to the two-sphere problem gives us an estimate of the value of radiative conductance one can expect from such an experiment. For microspheres with radii of 25 $\mu$m, we expect a radiative conductance between two silica spheres around 4.5 nWK$^{-1}$ at a gap of 200 nm. From Fig. \ref{fig:totcond10mic20mic}, we see that we can expect a conductance of 5.6 nWK$^{-1}$ at a gap of 100 nm. If the experimental configuration is not a two-sphere configuration but a sphere adjacent to a flat plate, the results of this chapter can be used as a guide to estimating the conductance. We expect trends to be similar -- that is we expect the increase in conductance to be of the form $Ax^{-n}$, where $x$ is the gap between the sphere and the flat plate. Most importantly, we expect $n$ to be a number between 0 and 1.


The work is supported by an ONR grant (Grant No. N00014-03-1-0835) through University of California, Berkeley.

\appendix
\section{\label{sec:appvsw}Properties of vector spherical waves}
\begin{equation}
\label{eqn:veccrosssim}
\oint_{\Omega} \left (\bm{V}_{lm}^{(s)}(\theta,\phi) \times \bm{V}_{pq}^{(s)*}(\theta,\phi) \right)\centerdot \bm{\hat{r}} d\Omega = 0
\end{equation}
\begin{equation}
\label{eqn:veccross23}
\oint_{\Omega} \left (\bm{V}_{lm}^{(2)}(\theta,\phi) \times \bm{V}_{pq}^{(3)*}(\theta,\phi) \right)\centerdot \bm{\hat{r}} d\Omega = \delta_{lp}\delta_{mq}
\end{equation}
\begin{equation}
\label{eqn:veccross32}
\oint_{\Omega} \left (\bm{V}_{lm}^{(3)}(\theta,\phi) \times \bm{V}_{pq}^{(2)*}(\theta,\phi) \right)\centerdot \bm{\hat{r}} d\Omega = -\delta_{lp}\delta_{mq}
\end{equation}
\begin{eqnarray}
\label{eqn:MMsource}
\nonumber k_f^2 \epsilon''_a\int_{V_a} \bm{M}_{lm}^{(1)}(k_a\bm{r'}_a) \centerdot \bm{M}_{pq}^{(1)*}(k_a\bm{r'}_a)  d\bm{r'} & = & \\
\delta_{lp}\delta_{mq}a\Im{\left(k_a^* a z_l^{(1)}(k_a a) \zeta_l^{(1)*}(k_a a)\right)} & &
\end{eqnarray}
\begin{eqnarray}
\label{eqn:NNsource}
\nonumber k_f^2 \epsilon''_a\int_{V_a} \bm{N}_{lm}^{(1)}(k_a\bm{r'}_a) \centerdot \bm{N}_{pq}^{(1)*}(k_a\bm{r'}_a)  d\bm{r'} & = & \\
\delta_{lp}\delta_{mq}a\Im{\left(k_a^* a z_l^{(1)*}(k_a a) \zeta_l^{(1)}(k_a a)\right)} & &
\end{eqnarray}

\newpage

\end{document}